\documentclass[12pt]{article}
\usepackage{epsfig}
\usepackage{amstex}
\usepackage{psfrag}
\def\F{{\cal F}}
\def\T{{\cal T}}

\def\dmu{\Delta\mu}

\title{Long range correlations in branched polymers.}
\author{Piotr Bialas\thanks{permanent address: Institute of Comp. Science, Jagellonian University, ul. Nawojki 11, 30-072 Krak\'ow, Poland}\\
{\em Universiteit van Amsterdam, Instituut voor Theoretische Fysica,}\\
{\em Valckenierstraat 65, 1018 XE Amsterdam,}\\
{\em The Netherlands}}
\date{}
\begin{document}
\maketitle
\begin{abstract}
We study the correlation functions in the branched polymer model.
Although there are no correlations in the grand canonical ensemble,
when looking at the canonical ensemble we find  negative long
range power like correlations. 
We propose that a similar mechanism explains the shape of recently measured
correlation functions in the elongated phase  of 4d simplicial gravity\cite{bj1}.
\end{abstract}

\section{Introduction}

Recently some measurements of the curvature--curvature correlation
functions in 4d simplicial gravity were presented \cite{bj1}. 
The data seems to
indicate  negative and power like behavior  both at the transition and
in the elongated phase. The latter seems surprising as one would expect
the exponential decay of correlations outside the critical region.
Those issues are hard to study numerically and even harder 
analytically. To gain some insight into possible mechanism of
those correlations we investigate here a simple geometrical model: 
branched polymers (BP). 

Besides its simplicity there are some other motivations for using this 
model. The BP describe the $c\rightarrow\infty$ limit of 2d simplicial
gravity \cite{a1,a2}. It is believed that BP also describe the elongated
phase of 4d simplicial gravity \cite{aj}. However the exact correspondence
between the quantities measured  in  ref. \cite{bj1} and BP is not known. 
Nevertheless
we can hope that the BP model captures the general features
of the elongated phase of 4d simplicial gravity.

\section{The model}

\begin{figure}
\begin{center}
\epsfig{file=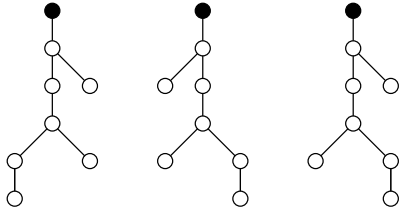}
\end{center}
\caption{\label{tree}Example of distinct planted rooted trees}
\end{figure}

\begin{figure}
\vskip5mm
\begin{center}
\epsfig{file=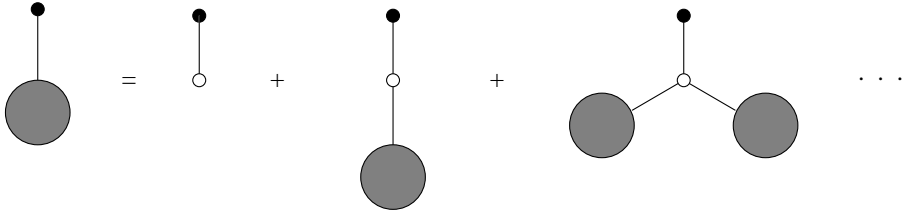,width=12cm}
\end{center}
\caption{\label{Zeqfig}Partition function}
\end{figure}

We consider here the ensemble of {\em planar rooted planted trees}. A
{\em tree} is a graph without the closed loops. A {\em rooted} tree is
a tree with one marked vertex (root) and {\em planted} tree is a tree
whose root's {\em degree} (number of branches) is one. Two trees are
considered as distinct if they cannot be mapped on each other by a
continuous deformation of the plane (see fig.~\ref{tree}).
We denote by
$n$ the total number of  non-root vertices in the tree and by $n_i$ the
number of non-root vertices of the degree $i$. Then the (grand)
partition function is defined as
\begin{equation}\label{Zdef}
Z(\mu,t)=\sum_{T\in\T}e^{-\mu n}\rho(T),\qquad \rho(T)=t_1^{n_1}t_2^{n_2}\cdots
\end{equation}
The $\T$ denotes the ensemble of all the trees. The 
equation for $Z(\mu,t)$  is (see fig.~\ref{Zeqfig}) \cite{a1} 
\begin{equation}\label{Zeq}
Z(\mu,t)=e^{-\mu}\F(Z(\mu,t))\quad\text{with}\quad 
\F(Z)=\sum_{i=1}^{\infty}t_iZ^{i-1}.
\end{equation}
The equation (\ref{Zeq}) can be rewritten as
\begin{equation}\label{Zeqbis}
e^{\mu}=\frac{\F(Z)}{Z}
\end{equation}
and the critical point $\mu_c$ corresponds to a value $Z_0$ where
the right hand side of (\ref{Zeqbis}) has a minimum. In the neighborhood
of this point
for a large class of parameters $t$ the partition function 
behaves like \cite{a1}
\begin{equation}\label{Zass}
Z(\mu,t)\approx Z_0(t) - Z_1(t) \sqrt{\dmu}+Z_2(t)\dmu+O(\dmu^{\frac{3}{2}})
\end{equation}  
with $\dmu=\mu-\mu_c$. That is the only class of solutions of the
equation (\ref{Zeq}) considered in this paper. Behaviour described by
(\ref{Zass}) is typical also for the elongated phase of the 4d simplicial
gravity \cite{aj}. From the equation (\ref{Zeq})
we can derive
\begin{equation}
\label{z0}
\begin{split}
Z_0&=\frac{\F(Z_0)}{\F^{(1)}(Z_0)},\\
\label{z1}
Z_1&=\sqrt{\frac{2\F(Z_0)}{\F^{(2)}(Z_0)}},\\
\label{z2}
Z_2&=\frac{1}{3}\frac{\F(Z_0)}{\F^{(2)}(Z_0)^2}(3Z_0\F^{(2)}-\F^{(3)}(Z_0)),
\end{split}
\end{equation}
where
$\F^{(i)}=\frac{\partial^i\F}{\partial Z^i}$.

\section{The Two-Point functions}

First we consider a ``volume--volume'' correlation function \cite{a2}
\begin{equation}
G(\mu,t;r)=\sum_{T\in\T_r}e^{-\mu n}\rho(T)
\end{equation}
where $\T_r$ is the ensemble of the trees with one  point marked at 
distance $r+1$ from the root. 

\begin{figure}
\psfrag{v1}{$v_1$}
\psfrag{vt}{$v_{r+1}$}
\begin{center}
\epsfig{file=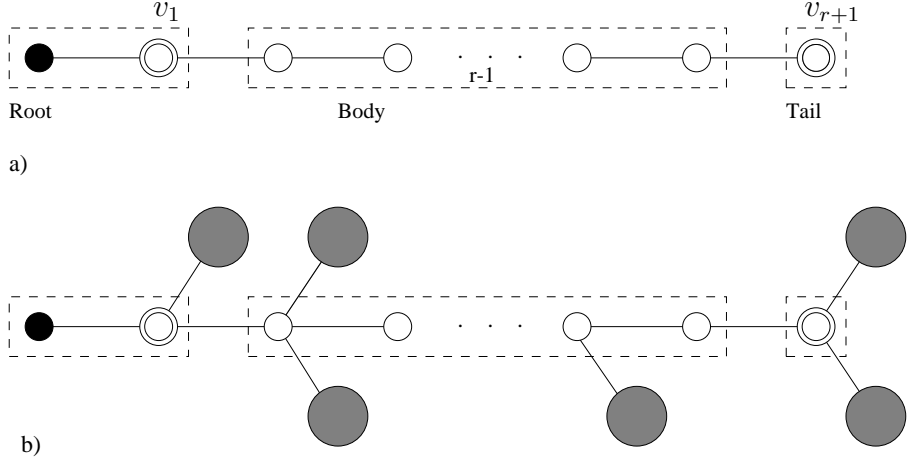,width=12cm,bbllx=0,bblly=60,bburx=428,bbury=295}
\end{center}
\caption{\label{Geqfig}Two-point function}
\end{figure}

The smallest possible tree in $\T_r$ is a chain of $r+1$ non-root
vertices which we split into root, body and tail (see
fig.~\ref{Geqfig}a).  The weight of this chain is
$t_1t_2^re^{-(r+1)\mu}$. All other configurations in $\T_r$ can be
obtained from this chain by attaching trees in its non-root
vertices(see fig.~\ref{Geqfig}b).  Attaching $n$ trees in the body or
root part of the chain corresponds to a factor $t_{n+2}(n+1)Z^n$. The
factor $n+1$ counts the possible relative positions of the chain.
Attaching $n$ trees to the tail corresponds to a factor $t_{n+1}Z^n$.
Finally our two-point function is
\begin{equation}\label{GZ}
\begin{split}
G(\mu,t;r)=&e^{-\mu}(t_2+2t_3Z+3t_4Z^2+\cdots)\cdot\\
&\cdot e^{-(r-1)\mu}(t_2+2t_3Z+3t_4Z^2+\cdots)^{r-1}e^{-\mu}(t_1+t_2Z+\cdots)\\
=&e^{-\mu}\F'(Z)e^{-(r-1)\mu}\F'(Z)^{(r-1)}e^{-\mu}\F(Z)
=(1+\frac{Z}{Z'})^rZ  
\end{split}
\end{equation}
where $\F'(Z)=\frac{\partial \F(Z)}{\partial Z}$ and
$Z'=\frac{\partial Z(\mu,t)}{\partial\mu}$.  In the last equality we
used the identity $e^{-\mu}\F'(Z)=1+\frac{Z}{Z'}$ obtained by
differentiating the equation (\ref{Zeq}) with respect to $\mu$.

We are interested in the Root-Tail correlation function defined
as follows:
\begin{equation}\label{RTdef}
RT(\mu,t;r)=\sum_{\T_r}d(v_1)d(v_{r+1})e^{-\mu n}\rho(T)
\end{equation}
where $d(v)$ is the degree of vertex $v$.  The form of this function
can be easily obtained by modifying $G(\mu,t;r)$
\begin{equation}
\label{RTform}
\begin{split}
RT(\mu,t;r)=&e^{-\mu}(2\cdot t_2+3\cdot2t_3Z+4\cdot3t_4Z^2+\cdots)\cdot\\
&\kern-10mm\cdot e^{-(r-1)\mu}(t_2+2t_3Z+3t_4Z^2+\cdots)^{r-1}e^{-\mu}(1\cdot t_1+2\cdot t_2Z+\cdots)\\
=&e^{-\mu}\frac{1}{Z}
\frac{\partial(\F'(Z)Z^2)}{\partial Z}e^{-(r-1)\mu}\F'(Z)^{(r-1)}e^{-\mu}
\frac{\partial(\F(Z)Z)}{\partial Z}\\
 =&(2+\frac{4Z}{Z'}+\frac{Z^2}{{Z'}^2}-\frac{Z^2Z''}{{Z'}^3})
(1+\frac{Z}{Z'})^{(r-1)}
(2Z+\frac{Z^2}{Z'})
\end{split}
\end{equation}
where we used the identity
$e^{-\mu}F''(Z)=\frac{2}{Z'}+\frac{Z}{{Z'}^2}- \frac{ZZ''}{{Z'}^3}$
again obtained by differentiating eq. (\ref{Zeq}) twice with respect
to $\mu$.

Similarly we define two other functions
\begin{equation}\label{Rdef}
\begin{split}
R(\mu,t;r)=
&\sum_{T\in\T_r}d(v_1)e^{-\mu n}\rho(T)
=(2+\frac{4Z}{Z'}+\frac{Z^2}{{Z'}^2}-\frac{Z^2Z''}{{Z'}^3})
(1+\frac{Z}{Z'})^{(r-1)}Z,
\end{split}
\end{equation}
\begin{equation}\label{Tdef}
\begin{split}
T(\mu,t;r)=&\sum_{T\in\T_r}d(v_{r+1})e^{-\mu n}\rho(T)
 =(1+\frac{Z}{Z'})^{r}
(2Z+\frac{Z^2}{Z'}).  
\end{split}
\end{equation}
Let 
\begin{equation}
\langle A \rangle_{\T_r} =\frac{\displaystyle\sum_{T\in\T_r}A(T)e^{-\mu n}\rho(T)}
{\displaystyle\sum_{T\in\T_r}e^{-\mu n}\rho(T)}.
\end{equation}
We  define the normalized connected Root-Tail correlation function as
\begin{equation}\label{RTc}
\begin{split}
RT_c(\mu,t;r)&=
\langle (d(v_1)-\langle d(v_1)\rangle_{\T_r})
(d(v_{r+1})-\langle d(v_{r+1})\rangle_{\T_r})\rangle_{\T_r}\\
&=\frac{RT(\mu,t;r)}{G(\mu,t;r)}-\frac{R(\mu,t;r)T(\mu,t;r)}{G(\mu,t;r)^2}.
\end{split}
\end{equation} 
It is easy to check that 
\begin{equation}
RT_c(\mu,t;r)\equiv0.
\end{equation}

\section{The canonical ensemble}

The functions defined in the preceding chapter are the discrete Laplace
transforms of their canonical counterparts e.g.
\begin{equation}\label{Glap}
G(\mu,t;r)=\sum_{n=0}^\infty e^{-\mu n}G(n,t;r)\quad\text{with}
\quad
G(n,t;r)=\sum_{T\in\T_r(n)}\rho(T)
\end{equation}
where $\T_r(n)$ is the ensemble of the trees belonging to $\T_r$ and
having exactly $n$ non-root vertices. To calculate the canonical
functions from grand canonical we have to perform the inverse of the
discrete Laplace transform. This usually can not be done
exactly and we proceed with a series of approximations.

The first approximation is that of replacing the discrete transform
(\ref{Glap}) with the continuous one. Then $G(n,t;r)$ is given by
the inverse Laplace transform which we calculate by the saddle
point method. The details are presented in appendix A.  Below we give
results to the leading order in $n$ ($r\ll n$).

\begin{equation}\label{Gn}
\begin{split}
G(n,t;r)\approx&\frac{1}{2\sqrt{\pi}}Z_0x
n^{-\frac{3}{2}}\,e^{\mu_c n}
\exp\left(-\frac{1}{4n}
x^2
\right),
\end{split}
\end{equation}
\begin{equation}\label{RTn}
\begin{split}
RT(n,t;r)\approx&\frac{2}{\sqrt{\pi}}
Z_0\left(x\frac{Z_0^2+Z_1^2}{Z_1^2}-\frac{Z_0(Z_0^2+6Z_0Z_2-5Z_1^2)}{Z_1^3}\right)
n^{-\frac{3}{2}}\,e^{\mu_cn}\cdot\\
&\cdot\exp\left(
-\frac{1}{4n}
\left(x-\frac{Z_0}{Z_1}
\frac{(Z_0^2+6Z_0Z_2-5Z_1^2)}{Z_0^2+Z_1^2}
\right)^2\right),
\end{split}
\end{equation}
\begin{equation}\label{Rn}
\begin{split}
R(n,t;r)\approx&\frac{1}{\sqrt{\pi}}Z_0\left(
x\frac{Z_1^2+Z_0^2}{Z_1^2}-\frac{Z_0(2Z_0^2+6Z_0Z_2-4Z_1^2)}{Z_1^3}\right)
n^{-\frac{3}{2}}\,e^{\mu_cn}\cdot\\
&\cdot\exp\left(
-\frac{1}{4n}\left(x-\frac{Z_0}{Z_1}
\frac{(2Z_0^2+6Z_0Z_2-4Z_1^2)}{Z_1^2+Z_0^2}
\right)^2\right),
\end{split}
\end{equation}
\begin{equation}\label{Tn}
\begin{split}
T(n,t;r)\approx&\frac{1}{\sqrt{\pi}}
Z_0\left(x+\frac{Z_0}{Z_1}\right)n^{-\frac{3}{2}}\,e^{\mu_cn}\exp\left(-\frac{1}{4n}\left(x+
\frac{Z_0}{Z_1}\right)^2\right)
\end{split}
\end{equation}
where we used $x=\frac{2rZ_0^2+Z_1^2}{Z_0Z_1}$. 

The connected correlation function $RT_c(n,t;r)$ is defined as in
(\ref{RTc}).
\begin{equation}\label{RTcn}
\begin{split}
RT_c(\mu,t;r)&=
\langle (d(v_1)-\langle d(v_1)\rangle_{\T_r(n)})
(d(v_{r+1})-\langle d(v_{r+1})\rangle_{\T_r(n)})\rangle_{\T_r(n)}\\
&=\frac{RT(n,t;r)}{G(n,t;r)}-\frac{R(n,t;r)T(n,t;r)}{G(n,t;r)^2}.
\end{split} 
\end{equation}
Before writing it down let us note that the terms proportional to
$x^2$ (and so to $r^2$) in the exponents cancel between numerators and
denominators in $RT_c(n,t;r)$. 
For large $n$ and $r/n\!\ll\!1$ we can neglect the
exponents.  The resulting expression is
\begin{equation}\label{RTf}
RT_c(n,t;r)\approx\frac{8Z_0^4(Z_0^2+3Z_0Z_2-2Z_1^2)}
{Z_1^2(2rZ_0^2+Z_1^2)}.
\end{equation}

\section{Examples and simulations}

Knowing the function $\F(Z)$ it is easy to calculate (\ref{RTf}). The
equation (\ref{z0}) can be solved numerically if the analytic solution
is not available. Below we give examples of two models. In both cases
the correlations are negative.  This was also the case for all other
models we tested and we are persuaded that this is a general feature
of the models with the expansion of the form (\ref{Zass}) and all the
weights $t$ positive.

\begin{enumerate}
\item[1)]
$t_1=t,\quad t_2=t_3=1,\quad t_4=t_5=\ldots=0$, 
$\F(Z)=t+Z+Z^2$, 
\begin{equation}\label{RTex1}
RT_c(n,t;r)=-\frac{4t}{(1+2\sqrt{t})(1+2\sqrt{t}+2r\sqrt{t})^2}.
\end{equation}
\item[2)]
$t_i=i^{-\frac{d}{2}}$, $\F(Z)=\frac{1}{Z}\text{Li}_{\frac{d}{2}}(Z)$\cite{a1}.
For $d=1$,
\begin{equation}\label{RTex2}
RT_c(n;r)=-\frac{1.3029}{(0.725958r+0.30219)^2}.
\end{equation}
\end{enumerate}

\begin{figure}[t]
\psfrag{infty}{$\scriptstyle\infty$}
\psfrag{xlabel}{$r$}
\psfrag{ylabel}[][][1][0]{$-RT_c(n,r)$}
\begin{center}
\epsfig{file=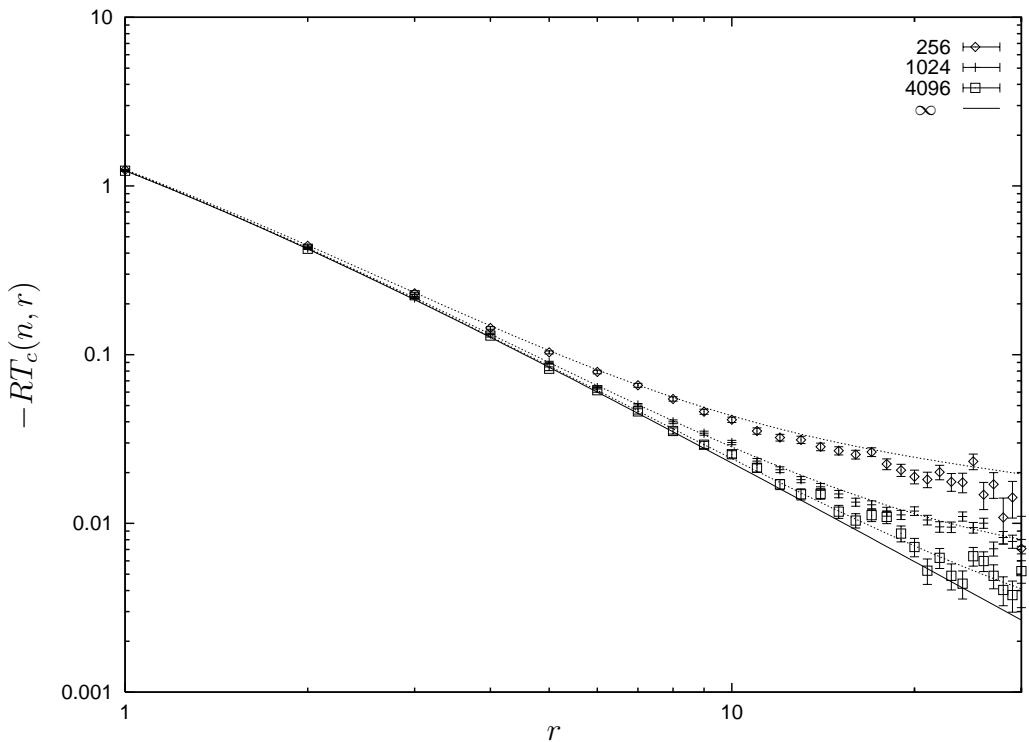,width=14cm}
\end{center}
\caption{\label{simul}The results of MC simulations.}
\end{figure}

The formula (\ref{RTf}) is valid for $n\rightarrow\infty$. To check
what finite size effects are to be expected we performed the MC
simulations of the second model ($d=1$) with $256$, $1024$ and $4096$
non-root vertices.  The results for the $RT_c$ are shown on
fig.~\ref{simul}.  We plotted the MC data, the large $n$ predictions
(formula (\ref{RTex2})) and the predictions for $RT_c$ without
neglecting the exponents (the dotted lines).

\section{Discussion}

\begin{figure}[h]
\psfrag{z1}{$w^2$}
\psfrag{z1n}{$0$}
\psfrag{z2}{$w^3$}
\psfrag{z2n}{$0$}
\psfrag{z3}{$wZ(w)$}
\psfrag{z3n}{$\frac{1}{9}Z(n)$}
\psfrag{z4}{$wZ(w)(1-w)-w^2$}
\psfrag{z4n}{$\frac{2}{9}Z(n)$}
\psfrag{z5}{$w^3Z(w)$}
\psfrag{z5n}{$\frac{1}{27}Z(n)$}
\psfrag{z6}{$w^Z(w)(1-w)-w^3$}
\psfrag{z6n}{$\frac{2}{27}Z(n)$}
\psfrag{z7}{$wZ(w)(1-2w)-w^2(1-w)$}
\psfrag{z7n}{$\frac{1}{9}Z(n)$}
\psfrag{z8}{$Z(w)(1-3w + w^2 + w^3)-w(1-2w)$}
\psfrag{z8n}{$\frac{4}{27}Z(n)$}
\hbox{\kern1cm\epsfig{file=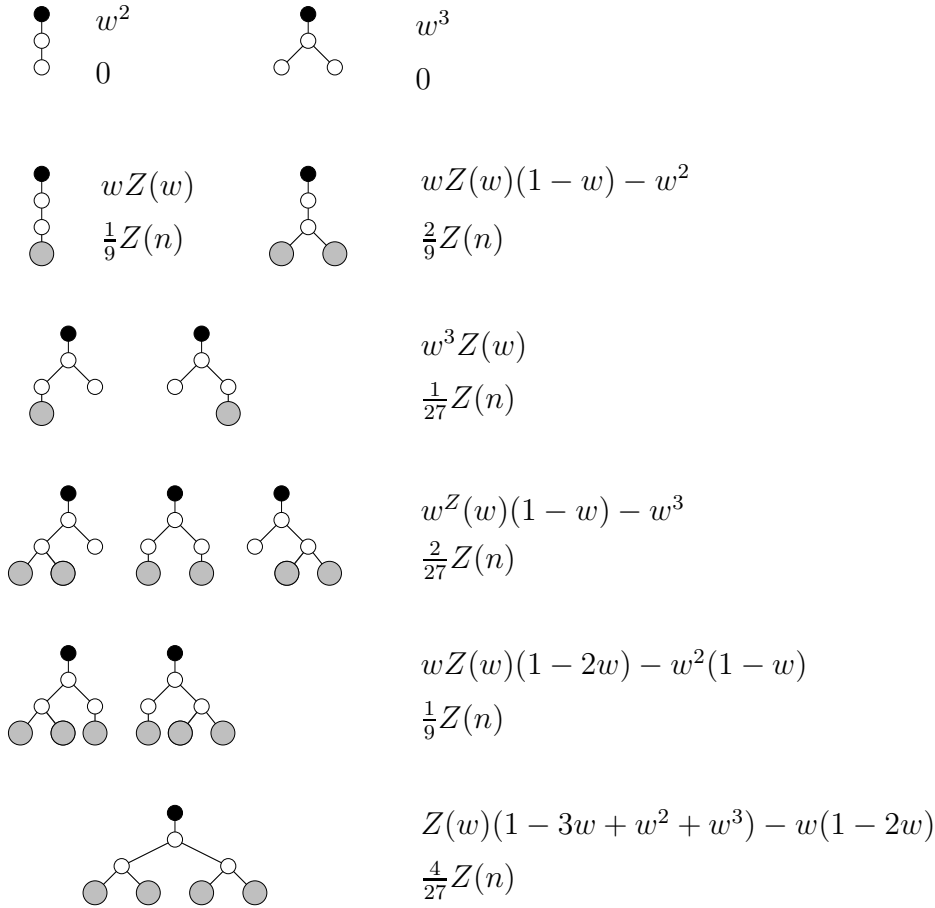,height=12cm}}
\caption{\label{corn}Calculation of $RT_c(r=1)$}
\end{figure}

The appearance of correlations in the canonical ensemble is not
surprising.  What is more surprising is that those correlations do not
vanish in the \hbox{$n\rightarrow\infty$} limit.  To understand better
what is happening we calculate $RT_c(r=1)$ for the first model from
previous section with \hbox{$t=1$} by the explicit tree counting. In
the fig.~\ref{corn} we list all the relevant groups of trees together
with their respective weights. The upper formulas refer to the grand
canonical ensemble and lower ones to canonical ensemble (valid for
$n\!\gg\!1$). Note that the first two trees cannot appear in
the canonical ensemble. If we calculate the $RT_c$ using the grand
canonical weights we get zero as expected. If we calculate the $RT_c$
with grand canonical weights but exclude from the sum two first trees
(those forbidden in canonical ensemble) we get a non-zero result
depending on the value of $w=\exp(-\mu)$. For the critical value
$w=\frac{1}{3}$ the result is $-\frac{9}{121}$.  Repeating the
calculations with canonical weights we obtain $-\frac{4}{75}$ which
agrees with (\ref{RTex1}) for $t=1$. This would indicate that the
effect is due to the absence of small configurations in the canonical
ensemble.

We have shown that a large class of BP models exhibits a long range
negative power like correlations in the canonical ensemble. Those
correlations are not finite size effects and survive in the
$n\rightarrow\infty$ limit. The shape of those correlations bears a
striking resemblance to the shape of correlation functions measured in
4d simplicial gravity \cite{bj1}. Before making a detailed comparison
one should  keep in mind that the exact correspondence between
the BP and 4d gravity is not known. The quantity measured in \cite{bj1}
has only a qualitative resemblance to the quantity calculated here. 
We believe however that the same mechanism can be responsible for
both.

\section{Acknowledgments}
I would like to thank Zdzislaw Burda, Jerzy Jurkiewicz and Jan Smit
for many helpful discussions and comments.  This work is supported by
the 'Stichting voor Fundamenteel Onderzoek der Matierie' (FOM). The
numerical simulations were partially carried out on the {\sc
PowerXplorer} at SARA.

\appendix

\section{Calculation of $G(n,t;r)$}

The inverse Laplace transform (continuous) of $G(\mu,t;r)$ is given by
\begin{equation}
G(n,t;r)\approx
\frac{1}{2\pi i}\int_{c-i\infty}^{c+i\infty}\text{d}\mu G(\mu,t;r)e^{n\mu}
\end{equation}
where $c>\mu_c$.  The saddle point equation is
\begin{equation}\label{sadle}
\frac{\partial\log(G(\mu,t;r))}{\partial \mu}=-n.
\end{equation}
Using the (\ref{Zass}) and (\ref{GZ}) we can expand the left hand side
of (\ref{sadle}).  Because we are interested in $n\gg1$ we keep only
the largest term of the expansion (in $\dmu$) of the left hand side of
(\ref{sadle}).  Solving this we obtain
\begin{align}
\mu_s\approx&
\frac{\left(r\frac{Z_0}{Z_1}+\frac{1}{2}\frac{Z_1}{Z_0}\right)^2}{n^2}+\mu_c,\\
\begin{split}
G(\mu_s,t;r)\approx&
Z_0\exp\left[-\frac{1}{n}\left(2\frac{Z_0^2}{Z_1^2}r^2
+2r+\frac{1}{2}\frac{Z_1^2}{Z_0^2}\right)\right],
\end{split}\\
\frac{\partial^2 \log(G(\mu_s,t;r))}{\partial \mu^2}\approx&
\frac{1}{2}\frac{n^3}{\left(r\frac{Z_0}{Z_1}+\frac{1}{2}
\frac{Z_1}{Z_0}\right)^2}.
\end{align}
Putting it all together we get the formula (\ref{Gn}).  Formulas
for other two-point functions can be obtained in a similar
way.

\end{document}